\documentstyle{article}
\title{Geometry of N=1 Super Yang-Mills Theory in Curved Superspace.}
\author{Anatoli Konechny and Albert Schwarz\\ Department of Mathematics, University of California, \\
Davis, CA 95616\\ KONECHNY@UCDMATH.UCDAVIS.EDU, \\
SCHWARZ@UCDMATH.UCDAVIS.EDU}
\date{June 2, 1996}
\begin{document}
\maketitle
\smallskip
\begin{abstract}
We give a new description of N=1 super Yang-Mills theory in curved superspace. 
It is based on the induced geometry approach to a curved superspace in which 
 it is viewed as a surface  embedded into ${\rm C}^{4|2}$.   
The complex structure on ${\rm C}^{4|2}$ supplied with a standard volume element 
induces a special Cauchy-Riemann (SCR)-structure on the embedded surface. 
We give an explicit construction 
 of SYM theory in terms of  intrinsic geometry of the superspace defined by this 
 SCR-structure and a CR-bundle over  the superspace. 
We write a manifestly SCR-covariant
 Lagrangian for SYM coupled with matter. We also show that in a special gauge 
our formulation coincides with   the standard one  which uses Lorentz connections. 
Some useful auxiliary results about the  integration over  surfaces in superspace 
 are  obtained. 

\end{abstract}
\section{Introduction} 
The main purpose of this work is to apply the induced geometry approach to N=1 
supergravity to the construction of N=1 SYM  theory on a curved superspace. 
This approach  was 
introduced in paper \cite{Gstr} .  We refer 
the reader to that paper for all details about the geometric constructions we 
use in the present one. However we give all necessary definitions and the paper 
can be read independently. The main benefit of induced geometry approach is 
that  it 
does not require any additional constraints like those one has to impose 
on a curvature and torsion tensors in the standard formulation of supergravity 
 (see \cite{BW} for a good exposition). The induced SCR-structure on a superspace 
incorporates all the necessary constraints and seems to be a more natural geometric 
construction then the Lorentz connections of the conventional approach. 
The induced geometry approach to N=1 supergravity was further developed in  papers 
\cite{gN1}, \cite{gN1 (2)}, \cite{auxdim}. In paper \cite{Rosly letter} the application to 
the construction of SYM theory over curved superspace was proposed. In the present 
work we  develop another point of view on this problem.

The paper is organized as follows. In section 2 we give some auxiliary results about 
the integration over  integral surfaces defined by $(0,2)$-dimensional distribution. 
 The derivation of these results is postponed until   appendix A.
Then we give the description of a curved superspace and CR-structure on it within 
the framework of induced  geometry approach. After that we explain how the results 
about the integration over integral surfaces can be applied to the construction of a
 chiral projecting operator.   
In section 3 we introduce some more geometric notions and define the gauge fields 
as sections of CR-bundles over the superspace. Then we formulate the main result 
and explain all the ingredients. The derivation of the Lagrangian is given in 
 appendix B. We finish this section by considering the proper reality 
conditions imposed on the fields used in the construction. As a result of these restrictions 
the set of fields reduces to the standard one. This procedure is similar to the one 
introduced in  \cite{Wessart}.   
 In section 4 we compare our constructions to the conventional ones. 
 
\section{Integration over integral surfaces determined by  
(0,2)-dimensional distributions }
Let $\cal M$ be a real $(m,n)$-dimensional supermanifold. 
Consider a pair of odd vector fields 
$E_{\alpha}$ (where $\alpha=1,2$) which are 
closed with respect to the anticommutator, i.e.
\begin{equation}    \label{anticom}
\{E_{\alpha},E_{\beta}\}=c_{\alpha\beta}^{\enspace \enspace \gamma}E_{\gamma}
\end {equation}
where $c_{\alpha\beta}^{\enspace \enspace \gamma}$ are some odd functions on
$\cal M$. By a super version of the Frobenius theorem,
 condition (\ref{anticom}) means that the vector fields $E_{\alpha}$
 determine an integrable distribution, i.e. for every point in  
$\cal M$ there exists a $(0,2)$-dimensional surface $\Sigma$  
going through it such that its tangent plane  at each point coincides 
with the one spanned  by $E_{\alpha}$ 
(equivalently our distribution defines a foliation which has the
surfaces $\Sigma$ as  leaves).
We are interested in the functions on $\cal M$ which are stable under
the action of the vector fields   $E_{\alpha}$, i.e. 
the functions $\Phi$ such that $E_{\alpha}\Phi=0$. Here and below we use
the same symbol for vector fields and for the first order differential
operators corresponding to them.

On every surface $\Sigma$ we have a natural volume element
defined by the requirement that the value of the volume form taken on the
fields $E_{1}$ and $E_{2}$ is equal to one.
Given an arbitrary function $\Phi$  one can get a function with
the property we want by integrating $\Phi$ over the leaves $\Sigma$. 
We will denote this operation by $\Box_{E} \Phi$ when applied to $\Phi$.
The explicit formula for $\Box_{E}$
 in terms of given
 $E_{\alpha}$ and functions $c_{\alpha\beta}^{\enspace \enspace \gamma}$ reads
 as
\begin{eqnarray} \label{mainformula}
\Box_{E}\Phi&=&
E^{\alpha}E_{\alpha}\Phi+c^{\alpha \enspace \sigma}_{\enspace 
\sigma}E_{\alpha}\Phi + \nonumber\\
& & +\Phi\left( \frac{2}{3}E^{\alpha}
c_{\alpha\sigma}^{\enspace \enspace \sigma}+
\frac{1}{3}c^{\alpha \enspace \sigma}_{\enspace \sigma}
c_{\alpha \beta}^{\enspace \enspace \beta} +
\frac{1}{6}c_{\alpha \beta}^{\enspace \enspace \sigma}
c_{\sigma}^{\enspace \alpha \beta} +
\frac{1}{12}c_{\alpha\beta\sigma}c^{\alpha\beta\sigma} \right) 
\end{eqnarray}
Here we are raising indices by means of the spinor metric tensor
$\epsilon^{\alpha\beta}$ ($\epsilon^{\alpha\beta}=
-\epsilon^{\beta\alpha}, \epsilon^{12}=1$) and lowering by means of
the inverse matrix $\epsilon_{\alpha\beta}$.
We give a  proof of this formula in  appendix A.
It is worth noting that the expression in parentheses in
formula (\ref{mainformula}) is just the volume of the leaf evaluated with
respect to the volume element specified above.

In order to get a function which is invariant with respect to a given
integrable distribution it is not necessary to use the natural volume element
related to the chosen pair of vector fields when integrating.
One can perform the integration  using any  volume element as well. 
But since it differs from the natural volume element by multiplication by
some function,  all possible freedom is reflected by the following formula
for a generic operation $\Box$:
\begin{equation} \label{freedom}
\Box\Phi = \Box_{\rho}\Phi\equiv \Box_{E}(\rho\Phi)
\end{equation}
where $\rho$ is some function. In short, this freedom is the same as modifying
the initial function by multiplying it by some fixed function.

Now let us consider the case of a complex supermanifold $\cal M$ with a pair of
complex vector fields $E_{\alpha}$ on it
satisfying the integrability condition (\ref{anticom}). In this case we have to
modify the consideration above slightly.
Formula (\ref{mainformula})  again defines an operation yielding a
$E_{\alpha}$-invariant function if one assumes that $E_{\alpha}$  is a
holomorphic vector field and $\Phi$ is a holomorphic function.
As before by the Frobenius theorem we have an integral complex
surface $\Sigma_{\mbox{\rm C}}$  going through every point in $\cal M$.
Our vector fields define a natural holomorphic volume element on 
$\Sigma_{\mbox{\rm C}}$. 
%%%%%%%%%%%%%%%
 For a generic real submanifold
$\Sigma_{\mbox{\rm R}}$, a real basis in a tangent space to 
$\Sigma_{\mbox{\rm R}}$ can be considered as a
complex basis in a tangent space to $\Sigma_{\mbox{\rm C}}$.
This means that the holomorphic volume form  determines a nondegenerate volume
element in a tangent space to $\Sigma_{\mbox{\rm R}}$.
Given such a generic submanifold
$\Sigma_{\mbox{\rm R}}\subset \Sigma_{\mbox{\rm C}}$, one can perform an 
integration of a holomorphic function over it. 
One can show that when we have a purely odd supermanifold $\Sigma_{\mbox{\rm C}}$, 
the result of integration does not  depend on the  particular choice
of $\Sigma_{\mbox{\rm R}}$. Therefore we see that in the case of a complex
manifold, the operation $\Box_{E_{\alpha}}$ 
(formula  (\ref{mainformula})) can be also interpreted in terms of integration
over the leaves.  

We are interested  in applications of formula (\ref{mainformula}) in the
framework of the induced geometry approach to the description of curved
superspace.  In this approach a curved superspace is described as a generic real
$(4,4)$-dimensional surface $\Omega$ embedded into ${\rm C}^{4|2}$
 (see \cite{Gstr} for details). The complex structure on  ${\rm C}^{4|2}$ 
induces a CR-structure on $\Omega$. This means that a  complex  plane is singled
out in every tangent space to $\Omega$.  
Namely if $T_{z}(\Omega)$ is the  real $(4,4)$-dimensional tangent space at a
point $z$ and $J$ denotes the linear operator given by the  multiplication 
by $i$, then $T_{z}(\Omega)\cap JT_{z}(\Omega)$ is the maximal complex subspace
contained in $T_{z}(\Omega)$. It is not difficult to figure out that  this
complex subspace is  of dimension $(0,2)$. One can choose vector fields 
\begin{equation}                 \label{vfields}
E_{\alpha}, {\bar E}_{\dot \alpha}, E_{c}\ 
(\alpha,\dot \alpha=1,2;c=1,\cdots,4)
\end{equation}
tangent to $\Omega$ such that the fields $E_{\alpha}$ form a (complex)
basis of the complex subspace at each point, 
the fields ${\bar E}_{\dot \alpha}$ define a basis in the complex conjugate 
plane   and the fields $E_{c}$ complete  $E_{\alpha}, {\bar E}_{\dot \alpha}$ 
to a (real)  basis of the whole tangent space. 
The (anti)commutator of two vector fields tangent to $\Omega$ is also a vector
field tangent to $\Omega$.  Thus we have 
\begin{equation} \label{C}
[E_{A},E_{B}\}=c_{AB}^{\enspace \enspace D}E_{D}
\end{equation}
where $c_{AB}^{\enspace \enspace D}$ are some functions and the indices take on
the values of the indices $\alpha, \dot \alpha, c$. The fact that our
CR-structure defined on $\Omega$ is induced by 
a $GL(4,2|{\rm C})$-structure in the ambient space implies 
\begin{equation} \label{intcr}
\{E_{\alpha},E_{\beta}\}=c_{\alpha \beta}^{\enspace \enspace \gamma}E_{\gamma}
\end{equation}
and the corresponding complex conjugate equations.
This means that we are dealing with an integrable CR-structure on $\Omega$. 
We call a function $\Phi$ defined on $\Omega$ chiral if 
$\bar E_{\dot \alpha}\Phi=0$. A function $\Phi^{+}$ is called antichiral if
$E_{\alpha}\Phi^{+}=0$. Note that the restriction to $\Omega$ of any holomorphic
function defined in some neighborhood of $\Omega$ in ${\rm C}^{4|2}$ 
is a chiral function (the converse is also true in some sense, see \cite{Gstr}).  
Formula (\ref{intcr}) looks exactly like formula (\ref{anticom}). The only
difference is that in the case at hand the complex fields $E_{\alpha}$ are
defined on a real manifold $\Omega$. Formula (\ref{mainformula}) can be used
to construct an (anti)chiral function from an arbitrary given one.
Moreover  one can give an interpretation of formula (\ref{mainformula}) similar
to those we gave for the purely real and complex cases using the 
complexification of  $\Omega$.

%%%%%%%%%%%%%%%%%%%%%%%%%%%%%%%%%%%%%%%%%%%%%%%%%%%%%%%%%%%
%%%%%%%%%%%%%%%%%          SECTION 2                     %%%%%%%%%%%%%%%%%%%%%%
%%%%%%%%%%%%%%%%%%%%%%%%%%%%%%%%%%%%%%%%%%%%%%%%%%%%%%%%%%%

\section{Formulation of N=1 SUYM in curved superspace
in terms of induced geometry and CR-bundles}
To construct the Lagrangian of N=1 super-Yang-Mills theory  in the induced 
geometry approach, first we need to say more about induced CR-structures and 
introduce some useful geometric notions.

Note that  the  basis (\ref{vfields}) of tangent vectors defining a CR-structure
on $\Omega$ is fixed up to  linear transformations  of the form
\begin{eqnarray} 
E'_{a}=g_{a}^{b}E_{b}+g_{a}^{\beta}E_{\beta}+\bar g_{a}^{\dot \beta}
\bar E_{\dot \beta} \nonumber \\
E'_{\alpha}=g_{\alpha}^{\beta}E_{\beta}, \, \, 
\bar E'_{\dot \alpha}=\bar g_{\dot \alpha}^{\dot \beta}\bar E_{\dot \beta} 
 \label{scrtr} \\
\end{eqnarray}
where $(g_{a}^{b})$ is a real matrix and $(\bar g_{\dot \alpha}^{\dot \beta})$
is the complex conjugate matrix of $(g_{\alpha}^{\beta}E_{\beta})$. 
If $\rm C^{4|2}$ is equipped with a volume element one can choose
$E_{a}, E_{\alpha}$ to be a unimodular complex basis in the  tangent space
to $\rm C^{4|2}$. This allows one to restrict the transformations (\ref{scrtr})
by the requirement   

\begin{equation} \label{scrdet}
\det(g_{a}^{b})=\det(g_{\alpha}^{\beta})
\end{equation}
In this case we say that there is an induced SCR-structure on $\Omega$.
>From now on we will assume that the
basis  (\ref{vfields})  defines a SCR-structure. 

For the functions $c_{AB}^{\enspace \enspace D}$ defined by formula (\ref{C})
in the case of induced SCR-structure we have the following identities
\begin{equation} \label{scrid}
c_{\alpha \beta}^{\enspace \enspace \dot \gamma}=c_{\alpha \beta}^{\enspace \enspace d}=0,\enspace
c_{\alpha \dot \beta}^{\enspace \enspace \dot \beta}=c_{\alpha d}^{\enspace \enspace d}
\end{equation}
and the  corresponding complex conjugate ones. 
%%%%%%%%%Levi form

We define the Levi matrix  of the surface $\Omega$ 
by the expression
$$ 
\Gamma^{a}_{b}=
i\bar \sigma^{\alpha \dot \beta}_{b}c_{\alpha \beta}^{a}
$$
where $\bar \sigma_{b}$ are the Pauli matrices for $b=1,2,3$ and the identity 
matrix for $b=0$.
The matrix $\Gamma^{a}_{b}$ coincides with the matrix of the Levi form defined
in the standard way (see \cite{Gstr}).   
 
%%%%%%%%%%%%%%%%%%%%%%%%%%%%%%%%%%%%%%%%%%%%%%

To construct a Yang-Mills theory on $\Omega \subset {\rm C}^{4|2}$, 
we start with two complex vector bundles $\cal F$ and ${\cal F}^{+}$
with structure group $G$.
Denote the Lie algebra corresponding to $G$ by $\cal G$.  
Trivializing  our bundles  we can represent their sections $\Phi$ and $\Phi^{+}$
locally as vector functions, called  fields.  
They  describe matter and charge conjugated matter respectively. Gauge
transformations correspond to the change of trivialization. They have the form
 \begin{equation} \label{gaugefi}
\Phi'=e^{i\Lambda}\Phi \qquad (\Phi^{+})'=e^{-i\bar \Lambda}\Phi^{+}
\end{equation}
where $\Lambda$ and $\bar \Lambda$ are some functions
(sections of corresponding homomorphism 
bundles)  with values in the representation of the Lie algebra $\cal G$
corresponding to the field $\Phi$. 
We want to stress the fact that for now we consider $\cal F$ and ${\cal F}^{+}$
separately, not requiring them to be complex conjugate bundles (the functions 
$\Lambda$, $\bar \Lambda$ in (\ref{gaugefi}) are also independent). 
By SUYM fields we understand two pairs of semiconnections
$$
\nabla_{\alpha}\Phi^{+} \equiv (E_{\alpha} + {\cal A}_{\alpha})\Phi^{+} \qquad 
\bar \nabla_{\dot \alpha}\Phi \equiv (\bar E_{\dot \alpha} + 
{\cal A}_{\dot \alpha})\Phi
$$
restricted by the conditions
\begin{equation} \label{zerocurv}
\{\nabla_{\alpha},\nabla_{\beta}\}=
c_{\alpha \beta}^{\enspace \enspace \gamma}\nabla_{\gamma}\qquad
\{\bar \nabla_{\dot \alpha},\bar \nabla_{\dot \beta}\}=
c_{\dot \alpha \dot \beta}^{\enspace \enspace \dot
\gamma}\bar \nabla_{\dot \gamma}
\end{equation}
These conditions mean that  the corresponding semiconnections have  vanishing
curvature.  It can be shown (see for example \cite{auxdim}) that  
semiconnections $\bar \nabla_{\dot \alpha}$ satisfying (\ref{zerocurv}) 
determine a CR-bundle  structure on ${\cal F}^{+}$, i.e.    ${\cal F}^{+}$
can be pasted together from trivial bundles by chiral gluing functions. 
One can define the chiral sections as those annihilated by 
$\bar \nabla_{\dot \alpha}$. Then condition (\ref{zerocurv}) guarantees that
there are sufficiently many of them. Analogously, given 
$\nabla_{\alpha}$ satisfying (\ref{zerocurv}), one obtains a
$\overline{\rm CR}$-bundle. By $\overline{\rm CR}$-bundle we mean a bundle whose
gluing functions are antichiral.  Moreover, one can take this property as a
definition of CR and $\overline{\rm CR}$-bundles (see \cite{auxdim} for
details). Thus the basic geometrical objects we start with are the surface 
$\Omega \subset {\rm C}^{4|2}$, the CR-bundle  $\cal F^{+}$ and the 
$\overline{\rm CR}$-bundle $\cal F$, both defined over $\Omega$.
The solutions to the zero curvature equations (\ref{zerocurv}) can be written
locally as  
\begin{equation} \label{UU}
{\cal A}_{\alpha}=e^{U}E_{\alpha}e^{-U} \qquad {\cal A}_{\dot \alpha}=
e^{-\tilde U}\bar E_{\dot \alpha}e^{\tilde U}
\end{equation}
where $U$ and $\tilde U$ are some $\cal G$-valued fields. 
Note that the fields $e^{-U}$ and $e^{\tilde U}$ are determined by (\ref{UU})
only up to the left multiplication by arbitrary antichiral and chiral fields
respectively. 
%Combining these transformations with the gauge ones defined by
%(\ref{gaugefi}) one can write the following general transformation for these  fields  
%\begin{equation} \label{trUU}
%e^{-U'}=e^{S^{+}}e^{-U}e^{-i\Lambda}\qquad e^{\tilde U'}=e^{T}e^{\tilde U}e^{i\bar \Lambda}
%\end{equation}
%where $T$ and $S^{+}$ are some chiral and antichiral functions respectively.
%Remind that in flat space the Lagrangian of SYM theory has the form 
%\begin{eqnarray} 
%\cal L_{fl}&=&\int d\theta d\theta \frac{1}{4k}W^{\alpha}W_{\alpha}  + 
%\int d\theta d\theta d\bar \theta d\bar \theta \Phi^{+}e^{V}\Phi  +  \nonumber \\
%&& +  \int d\theta d\theta \frac{1}{2}m_{ij}\Phi_{i}\Phi_{j} + \frac{1}{3}g_{ijk}\Phi_{i}\Phi_{j}\Phi_{k} 
%+ h.c.  \label{flatym}
%\end{eqnarray}
If we want to write a gauge invariant Lagrange function
for chiral fields  we will immediately encounter the difficulty in writing the
kinetic term, which for the case of free chiral fields 
is simply $\Phi^{+}\Phi$. This difficulty is due to the fact that in the case
at hand these fields are  sections of different bundles. Therefore  we are
forced to identify CR and $\overline{\rm CR}$ bundles choosing a section  of 
the bundle ${\cal F}^{+}\otimes{\cal F}^{*}$. 
Here ${\cal F}^{*}$ is the dual to the bundle ${\cal F}$. This section we denote by
$e^{V}$. Under the gauge transformations the field $e^{V}$ transforms in the
following way
\begin{equation} \label{gaugeV}
e^{V'}=e^{-i\Lambda^{t}}e^{V}e^{i\bar \Lambda}
\end{equation}
Now we can take the gauge invariant combination $\Phi^{+}e^{V}\Phi$ as a kinetic
density term.

Our next goal is to describe a gauge invariant theory in terms of the fields 
$\Phi_{i}, \Phi_{i}^{+}, e^{-U}, e^{\tilde U}, e^{V}$ defined on our curved
superspace $\Omega$.     
This will be done  in a manifestly SCR-covariant way, i.e. independently of the
choice of basis vector fields (\ref{vfields}) up to local SCR-transformations
(\ref{scrtr}). Instead of the customary Lorentz connections in our construction
of the Lagrangian, we use only objects defined by the internal geometry of the
superspace $\Omega$, namely the Levi matrix $\Gamma$ and the functions 
$c_{AB}^{\enspace \enspace D}$. 
We postpone the details of this construction until  appendix B.  
Now we want to formulate the main result. The Lagrangian  has the following form 
\begin{eqnarray}
{\cal S}&=&  
\int dV\left[\frac{1}{k}(\det\Gamma)^{-1}(\Box_{\bar {\cal D}}e^{G}E^{\alpha}
e^{-G}) (\Box_{\bar {\cal D}}e^{G}E_{\alpha}e^{-G})\right] + \nonumber \\
&&+ \int dV\left[\bar \Box_{\bar E}\left| \frac{1}{4}\det\Gamma
\right|^{-\frac{1}{3}}\Phi^{+}e^{V}\Phi\right] + \nonumber \\
&& + \int dV\left[a_{i}\Phi_{i}+\frac{1}{2}m_{ij}\Phi_{i}\Phi_{j}+
\frac{1}{3}g_{ijk}\Phi_{i}\Phi_{j}\Phi_{k} \right] + h.c. \label{scrlag}
\end{eqnarray}

Here as in flat space the N=1 SUYM Lagrangian contains a Lagrangian of gauge
fields, a kinetic term of chiral fields and a term describing
the interaction between  chiral fields. 
In (\ref{scrlag}) we  are using the following notations: $k$ is a coupling
constant, $e^{G}=e^{\tilde U}e^{-V}e^{-U^{t}}$, $a_{i}, m_{ij}, g_{ijk}$ are
coupling constants which must be chosen in a way that ensures the gauge
invariance of matter-matter interaction, $\Box_{\bar E}$ is a chiral projector
whose general form was described in section 2, and $\Box_{\bar {\cal D}}$ is a
"covariant" chiral projector which is constructed as $\Box_{\bar E}$  
with  derivatives $\bar E_{\alpha}$ replaced by "covariant" derivatives
$\bar {\cal D}_{\dot \alpha}$. More precisely, $\bar {\cal D}_{\dot \alpha}$
acts on an arbitrary tensor $V$ carrying the spinor index $\alpha$ in the
following way
$$
{\bar {\cal D}_{\dot \alpha}}V_{\beta} = \bar E_{\dot \alpha}V_{\beta} - 
{\check c}_{\dot \alpha, \dot \sigma \beta}^{\enspace \enspace \enspace
\, \dot \sigma \sigma} V_{\sigma} 
$$ 
where ${\check c}_{\dot \alpha, \dot \sigma \beta}^{\enspace \enspace
\enspace \, \dot \sigma \sigma}$
are some functions which can be expressed in terms of ${c_{AB}}^{D}$ 
(see formulae (\ref{crel}), (\ref{twobas}), (\ref{a}) in  appendix B).
Note that the integrands in (\ref{scrlag}) are chiral functions (for the
first term this fact follows from its construction, which is explained in
details in appendix B). 
The integration in (\ref{scrlag}) should be understood as a chiral integration.
%%%Anatoli, the following expression is not a sentence.
If a chiral function can be extended to a holomorphic function in some domain in
$\rm C^{4|2}$ where it can be integrated with the holomorphic volume element
over a real (4,2)-submanifold contained in $\Omega$. 

%%%%%%%%%%%%%%%%%%%%%%%%%%%%%%%%%%%%%%%%%%%%%%%%%%%%%%%
%%%%%%%%%%%%%%%%            REALITY CONDITIONS       %%%%%%%%%%%%%%%%%%%%%%%%%%
%%%%%%%%%%%%%%%%%%%%%%%%%%%%%%%%%%%%%%%%%%%%%%%%%%%%%%%%%%%%%

We have constructed a Lagrangian depending on the field $V$, 
the specific combination of the fields
 $e^{U}, e^{\tilde U}, e^{V}$ which we denoted by $e^G$ , and matter fields
 $\Phi_{i}, \Phi_{i}^{+}$.  All these fields are complex. In order to perform 
a functional integration over the fields $U, \tilde U, V, \Phi_{i}$ one has to
restrict these fields to a real surface in functional space. Once this surface
is chosen this will restrict our large gauge group to a smaller one. 
The real surface we choose is given by the equations 
\begin{equation} \label{realcond}
 \bar \Phi_{i}=\Phi_{i}^{+}  , \, V=\bar V^{t} , \,  -\bar U=\tilde U
\end{equation}  
where the upper bar denotes complex conjugation.
The gauge transformations preserving these reality conditions are of the form
(\ref{gaugefi}), (\ref{gaugeV}) where $\Lambda$ and $\bar \Lambda$ are complex
conjugates of each other and $T=\bar S^{+}\equiv S $. Still we have a rather
large gauge group. Let us do a partial gauge fixing by requiring that $e^{U}=1$.
By reality conditions (\ref{realcond}) this implies $e^{-\tilde U}=1$ and 
therefore $e^{-G}=e^{V}$, which means that our Lagrangian  (\ref{scrlag})
contains only the field $V$ in this partial gauge fixing. The remaining gauge
group contains the transformations with $i\Lambda=S^{+}, \, -i\bar \Lambda=S$,
i.e. the antichiral transformations of the fields $\Phi_{i}$ 
and the corresponding complex conjugate chiral transformations of the fields
$\Phi_{i}^{+}$. 
  
%%%%%%%%%%%%%%%%%%%%%%%%%%%%%%%%%%%%%%%%%%%%%%%%%%%%%
%%%%%%%%%%%%%%%                 SECTION 3      %%%%%%%%%%%%%%%%%%%%%%%
%%%%%%%%%%%%%%%%%%%%%%%%%%%%%%%%%%%%%%%%%%%%%%%%%%%%%

\section{Comparison with the conventional  approach} 
In this section we want to compare our constructions with the conventional
(Wess-Zumino) approach to supergravity 
and super Yang-Mills theory in curved superspace which is presented in \cite{BW} 
in great detail. 
We start with a comparison of formula (\ref{mainformula}) with the standard
formula for (anti)chiral projection operators  (see \cite{BW}, chapter 19). 
But first let us recall briefly the main constituents of the conventional
approach.  
%%%%%%%%%%%%%%%pure Bagger-Wess
 In this approach we have a connection defined on a tangent bundle over
 $(4,4)$-dimensional real superspace with the Lorentz group as structure group.
Another dynamical variable in this approach is a vielbein $E^{M}_{A}$ which
defines the covariant derivatives ${\cal D}_{M}$ on the tangent bundle and also
identifies the Lorentz bundle with the tangent one allowing to transform 
world indices into Lorentz indices and vice-versa. Here we are working
with Lorentz indices.  Under a certain set of constraints
(see \cite{BW}) the vector fields 
$E_{\alpha}=E_{\alpha}^{M}\partial_{M}$ are closed with respect to the 
anticommutator.  Moreover the functions 
$c_{\alpha\beta}^{\enspace  \enspace \gamma}$ defining the anticommutation
relations satisfy the following condition

\begin{equation} \label{connect}
c_{\alpha\beta}^{\enspace  \enspace \gamma}=
\omega_{\alpha\beta}^{\enspace  \enspace \gamma}+
\omega_{\beta\alpha}^{\enspace  \enspace \gamma}
\end{equation} 
where  $\omega_{\alpha\beta}^{\enspace  \enspace \gamma}$ are the connection
coefficients for the covariant differentiation of spinor fields, having only
Lorentz indices.

The antichiral projector acts on arbitrary function $\Phi$ as follows
\begin{equation} \label{8R}
({\cal D}^{\alpha}{\cal D}_{\alpha} - 8R^{+})\Phi
\end{equation}
and gives an antichiral function as a result. Here 
$R^{+}=\frac{1}{24}R_{\alpha\beta}^{\enspace \enspace \alpha \beta}$
denotes the invariant  obtained from the curvature tensor 
$$
R_{\alpha\beta\delta}^{\enspace \enspace \enspace \gamma}=
E_{\alpha}\omega_{\beta\delta}^{\enspace\enspace\gamma}
+E_{\beta}\omega_{\alpha\delta}^{\enspace\enspace\gamma}+
\omega_{\alpha\sigma}^{\enspace\enspace\gamma}
\omega_{\beta\delta}^{\enspace\enspace\sigma}+
\omega_{\beta\sigma}^{\enspace\enspace\gamma}
\omega_{\alpha\delta}^{\enspace\enspace\sigma}-
c_{\alpha\beta}^{\enspace\enspace\sigma}
\omega_{\sigma\delta}^{\enspace\enspace\gamma}
$$
After the contraction of indices this gives the following expression for
$-8R^{+}$
\begin{equation} \label{R(omega)}
-8R^{+}=-(E_{\alpha}\omega_{\beta}^{\enspace\alpha\beta}+
E_{\beta}\omega_{\alpha}^{\enspace\alpha\beta}+
\omega_{\beta\sigma}^{\enspace\enspace\beta}
\omega_{\alpha}^{\enspace\alpha\sigma})+
\omega^{\delta\alpha\beta}\omega_{\alpha\beta\delta}
\end{equation}
%%%%%%%pure Begger-Wess finished
The induced geometry approach to supergravity has been shown to be equivalent
to the Wess-Zumino one (\cite{Gstr}). Thus one can use formula 
(\ref{mainformula}) to obtain (\ref{8R}). 
We identify the pair of vector fields $E_{\alpha}$ appearing  in the Wess-Zumino
approach with those in definition of the $\overline{\rm CR}$-structure induced on 
$\Omega$ (i.e. complex conjugate to the corresponding CR-structure).  
 The formulae (\ref{8R}) and
(\ref{mainformula}) must be equivalent  at least  up to the freedom described by
formula (\ref{freedom}). Indeed as one can easily check, substituting relation
(\ref{connect}) in (\ref{mainformula}), we will get exactly formula (\ref{8R})
with the term $-8R^{+}$ expressed as in (\ref{R(omega)}).

If one starts with the induced geometry approach then in order to get the Lorentz
gauge group one has to require the following gauge condition 
$$
{c_{\alpha, \dot \beta}}^{a}=2i\sigma_{\alpha \dot \beta}^{a} 
$$ 
where  $\sigma_{\alpha \dot \beta}^{a} $ are the Pauli matrices.
This condition fixes the SCR-basis 
(\ref{vfields}) up to transformations of the form (\ref{scrtr}) where 
$\det(g_{a}^{b})=\det(g_{\alpha}^{\beta})=
\det(\bar g_{\dot \alpha}^{\dot \beta})=1$, i.e. up to the Lorentz 
transformations. In this gauge $4(\det\Gamma)^{-1}=1$, which as it is shown in
appendix B (see formulae (\ref{important}) and (\ref{conjimp})) implies 
$$
{{\check c}_{\gamma, \sigma \dot \sigma}}^{\enspace \enspace \enspace \,
\sigma \dot \sigma} = 
{{\check c}_{\dot \gamma, \dot \sigma  \sigma}}^{\enspace \enspace \enspace
\, \dot \sigma  \sigma} = 0 
$$ 
Moreover, the quantities 
$-{\check c}_{\dot \alpha, \dot \sigma \beta}^{\enspace
\enspace \enspace \, \dot \sigma \sigma}, 
-{\check c}_{ \alpha,  \sigma \dot \beta}^{\enspace \enspace \enspace \,
\sigma \dot \sigma} $ 
transform now as coefficients of the Lorentz connection (as can be seen from
(\ref{trcheckc}) for Lorentz transformations). Thus it seems reasonable to
identify  $-{\check c}_{\dot \alpha, \dot \sigma \beta}^{\enspace
 \enspace \enspace \, \dot \sigma \sigma}$ and  $
-{\check c}_{ \alpha,  \sigma \dot \beta}^{\enspace \enspace \enspace \,
\sigma \dot \sigma}$ 
with the connection coefficients ${\omega_{\dot \alpha}}^{\beta}_{\sigma}$ and
${\omega_{\alpha}}_{\dot \beta}^{\dot \sigma}$ respectively, from the
conventional approach. The last assumption implies 
$$ 
\Box_{\bar {\cal D}}V_{\alpha} =
(\bar {\cal D}_{\dot \gamma}\bar {\cal D}^{\dot \gamma} - 
8R)V_{\alpha} 
$$ 
and the corresponding complex conjugate identity. This shows that in the gauge
specified above, the  Yang-Mills Lagrangian term 
from (\ref{scrlag}) reduces to the standard one, which is written in terms of
field strengths $\bar W_{\dot \alpha}$.

Indeed, the identities $-{\check c}_{\dot \alpha, \dot
\sigma \beta}^{\enspace \enspace \enspace \, \dot \sigma \sigma} 
= {\omega_{\dot \alpha}}^{\beta}_{\sigma}, 
-{\check c}_{ \alpha,  \sigma \dot \beta}^{\enspace \enspace
\enspace \,  \sigma \dot \sigma} =  
{\omega_{\alpha}}_{\dot \beta}^{\dot \sigma}$ 
are true. One can derive them from the standard set of torsion constraints

\begin{eqnarray}
&&T_{\underline{\alpha} \underline{\beta} }^{\underline{\gamma}} = 0, \, 
T_{\alpha \beta}^{a}=T_{\dot \alpha \dot \beta}^{a} = 0
\nonumber \\
&& T_{\alpha \dot \beta}^{a}=2i\sigma_{\alpha \dot \beta}^{a}, \,
T_{\underline{\alpha} b}^{a}=0 , \, 
T_{ab}^{c}=0
\end{eqnarray}
where $\underline{\alpha}$ denotes either $\alpha$ or $\dot \alpha$.
Finally, the volume element $dV$ we used in (\ref{scrlag}) is nothing but the
chiral volume element of the conventional approach, usually 
denoted as ${\cal E}d^{2}\Theta$. This completes the derivation of the
conventional picture from the induced geometry one.

%%%%%%%%%%%%%%%%%%%%%%%%%%%%%%%%%%%%%%%%%%%%%%%%%%%%
%%%%%%%%%%%%%%%%%           APPENDICES                %%%%%%%%%%%%%%%%%%
%%%%%%%%%%%%%%%%%%%%%%%%%%%%%%%%%%%%%%%%%%%%%%%%%
\appendix
\section{Explicit formula for the chiral projector}

To prove  formula (\ref{mainformula}) let us first consider the case of
anticommuting vector fields, i.e. the functions
$c_{\alpha\beta}^{\enspace \enspace \gamma}$ are identically zero.

Let $X ^{M}$ be coordinates on $\cal M$ ($M=1,\cdots, m+n$),
so we have  $E_{\alpha}=E_{\alpha}^{M}\partial_{M}$
where $\partial_{M}$ is the derivative with respect to $X^{M}$ and
$E_{\alpha}^{M}$ are the coordinates of the field $E_{\alpha}$.  
et the surface $\Sigma$ be defined in a parametric form $X^{M}=X^{M}(\xi)$
where $\xi=(\xi^{\alpha}), (\alpha=1,2$) are odd coordinates on  $\Sigma$. 
Then we have the following equation for  $\Sigma$
\begin{equation} \label{sig}
\frac{\partial X^{M}(\xi )}{\partial\xi^{\alpha}}=E_{\alpha}^{M}(X(\xi))
\end{equation}
Since we assume the fields $E_{\alpha}$ anticommute we can integrate this
equation expanding $X^{M}(X_{0},\xi)$
in $\xi$ (here $X_{0}$ is an initial data for the system (\ref{sig}) ).
The solution to (\ref{sig}) reads as follows
\begin{equation} \label{sigeq}
X^{M}(X_{0},\xi )=X^{M}_{0}+\xi^{\alpha}E_{\alpha}^{M}+
\frac{1}{4}\xi\xi E^{\sigma}(E_{\sigma}^{M})
\end{equation} 
where $E_{\alpha}^{M}$ and $E^{\sigma}(E_{\sigma}^{M})$ are evaluated at the
point $\xi=0$ and $\xi\xi=\xi^{\alpha}\xi_{\alpha}$. 
The natural volume element on $\Sigma$ in $\xi$-coordinates is simply
$d\xi d\xi$.  Expanding the given function
$\Phi$ in $\xi$ and performing the integration we get
\begin{equation} \label{EE}
(\Box_{E}\Phi)(X_{0})=\int \Phi(X(X_{0},\xi )) d\xi d\xi =
E^{\alpha}E_{\alpha}(\Phi)
\end{equation}
which is obviously  annihilated by $\tilde E_{\alpha}$.
As one can easely see formula (\ref{EE}) coincides with (\ref{mainformula})
when $c_{\alpha\beta}^{\enspace \enspace \gamma}=0$.
Thus for the case of anticommuting vector fields
formula (\ref{mainformula}) is proved.

Now let us consider two different pairs of vector fields 
$E_{\alpha}$ and ${\tilde E}_{\alpha}$ defining the same integrable
distribution.
Let $c_{\alpha\beta}^{\enspace \enspace \gamma}$ and 
${\tilde c}_{\alpha\beta}^{\enspace \enspace \gamma}$ be
the corresponding functions defining the anticommutation relations for each
pair. The fields $E_{\alpha}$ and ${\tilde E}_{\alpha}$  are connected by means
of some nondegenerate even matrix $A$:
\begin{equation} \label{transform}
 {\tilde E}_{\alpha}=A_{\alpha}^{\beta}E_{\beta} 
\end{equation} 
The transformation law for the functions 
$c_{\alpha\beta}^{\enspace \enspace \gamma}$ reads as follows

\begin{equation} \label{ctransf}
{\tilde c}_{\alpha\beta}^{\enspace \enspace \gamma}=
A_{\alpha}^{\delta}A_{\beta}^{\sigma}(A^{-1})^{\gamma}_{\mu}
c_{\delta \sigma}^{\enspace \enspace \mu} +
(A^{-1})^{\gamma}_{\delta}(A_{\alpha}^{\sigma}
E_{\sigma}A_{\beta}^{\delta}
+ A^{\sigma}_{\beta}E_{\sigma}A_{\alpha}^{\delta}) 
\end{equation}

Since the ratio of the natural supervolume element corresponding to the
vector fields ${\tilde E}_{\alpha}$ to the one of 
the fields  $E_{\alpha}$ equals $\det A$,
the operations $\Box_{E}$ and $\Box_{\tilde E}$
are related by the following formula
\begin{equation} \label{consq}
 \Box_{\tilde E}\Phi=\Box_{E}(\det A\Phi)    
\end{equation}

Note that since our distribution is integrable, given the vector fields
$E_{\alpha}$ we can perform 
a linear transformation (\ref{transform})  
such that  the new vector fields  ${\tilde E}_{\alpha}$ will anticommute.
Thus if we prove that the right hand side of (\ref{mainformula}) under
the transformations (\ref{transform}) satisfies (\ref{consq}) then 
we  will have proved formula  (\ref{mainformula}).
We are going to show this for the case of infinitesimal transformations
$A^{\beta}_{\alpha}=\delta_{\alpha}^{\beta}+
a^{\beta}_{\alpha}$, where $a^{\beta}_{\alpha}$ is an infinitesimally small 
matrix. But this also proves (\ref{mainformula}) because
every finite transformation (\ref{transform}) can be obtained as a succession
of infinitesimally small ones.

Keeping only the terms of the first order in $a$, we have the following
transformations

\begin{eqnarray}            \label{inftesim}
{\tilde E}_{\alpha}&=&E_{\alpha}+a^{\beta}_{\alpha}E_{\beta}
\nonumber\\
{\tilde E}^{\alpha}&=&E^{\alpha}-{\bar a}_{\beta}^{\alpha}E^{\beta} \\
{\tilde c}_{\alpha\beta}^{\enspace \enspace \gamma}&=&
c_{\alpha\beta}^{\enspace \enspace \gamma}+
a_{\alpha}^{\sigma}c_{\sigma \beta}^{\enspace \enspace \gamma}+
a_{\beta}^{\sigma}c_{\sigma \alpha}^{\enspace \enspace \gamma} -                      
a^{\gamma}_{\sigma}c_{\alpha \beta}^{\enspace \enspace \sigma} +
E_{\alpha}a_{\beta}^{\gamma} + E_{\beta}a_{\alpha}^{\gamma} \nonumber
\end{eqnarray}
where 
${\bar a}^{\beta}_{\alpha}=
\epsilon_{\alpha \nu}\epsilon^{\beta \mu}a^{\nu}_{\mu}$. 
Note that 
$a_{\alpha}^{\beta}-{\bar a}^{\beta}_{\alpha}=
\delta^{\beta}_{\alpha}tr(a)$
where $\delta^{\beta}_{\alpha}$ is the Kronecker symbol. Another useful
identity is 
$$
E^{\delta}E_{\sigma}a^{\sigma}_{\delta}=
\frac{1}{2}E^{\delta}E_{\delta}(tr(a)) + 
\frac{1}{2}c_{\enspace \sigma}^{\delta \enspace \gamma}
E_{\gamma}a^{\sigma}_{\delta}
$$
For the bilinear combinations in  $c_{\alpha\beta}^{\enspace \enspace \gamma}$
which enter (\ref{mainformula}) and for the operator $E^{\alpha}E_{\alpha}$,
up to the first order in $a$ we have
\begin{eqnarray} \label{quadrtr}
{\tilde c}^{\alpha \enspace \sigma}_{\enspace \sigma}
{\tilde c}_{\alpha \beta}^{\enspace \enspace \beta}&=&
(1+tr(a))c^{\alpha \enspace \sigma}_{\enspace \sigma}
c_{\alpha \beta}^{\enspace \enspace \beta}+
2c^{\alpha \enspace \sigma}_{\enspace \sigma}E_{\alpha}a^{\beta}_{\beta} +
2c^{\alpha \enspace \sigma}_{\enspace \sigma}E_{\beta}a^{\beta}_{\alpha}
 \nonumber\\
{\tilde c}_{\alpha \beta}^{\enspace \enspace \sigma}
{\tilde c}_{\sigma}^{\enspace \alpha \beta}&=&
(1+tr(a))c_{\alpha \beta}^{\enspace \enspace \sigma}
c_{\sigma}^{\enspace \alpha \beta}+
2c_{\sigma\alpha}^{\enspace \enspace \beta}E^{\alpha}a_{\beta}^{\sigma} - 
2c_{\sigma}^{\enspace \alpha \beta}E_{\beta}a_{\alpha}^{\sigma} \\
{\tilde c}_{\alpha\beta\sigma}{\tilde c}^{\alpha\beta\sigma}&=&(1+tr(a))
c_{\alpha\beta\sigma}c^{\alpha\beta\sigma}+ 
4c^{\alpha\beta}_{\enspace \enspace \sigma}E_{\alpha}a_{\beta}^{\sigma} 
\nonumber\\
{\tilde E}^{\alpha}{\tilde E}_{\alpha}&=&(1+tr(a))E^{\alpha}E_{\alpha} +
(E^{\alpha}a_{\alpha}^{\beta})E_{\beta} \nonumber
\end{eqnarray} 

Note that the appearance of the combination $1+tr(a)$ in (\ref{quadrtr})
is natural because for the volume preserving transformations
(\ref{transform}) the contraction of upper and lower indices becomes an
invariant operation with respect to the derivative independent
part of transformation (\ref{ctransf}).
Substituting the expressions (\ref{inftesim}) and (\ref{quadrtr}) in the
right hand side of (\ref{mainformula}) using the identities mentioned above and
keeping terms only up to the first order in $a$,
we obtain the following expression

\begin{eqnarray}  
(1+tr(a))E^{\alpha}E_{\alpha}\Phi + ((1+tr(a))
c^{\alpha \enspace \sigma}_{\enspace  \sigma}+ 
2(E^{\alpha}tr(a)))E_{\alpha}\Phi + 
\nonumber\\ + \Phi(E^{\alpha}E_{\alpha}tr(a) +
c^{\delta \enspace \sigma}_{\enspace \sigma}E_{\delta}tr(a)
+ (1+tr(a)){\cal V}(c))   \label{b}
\end{eqnarray}
where for compactness of notation we denoted the term
multiplying $\Phi$ in (\ref{mainformula}) by ${\cal V}(c)$.
Now after some trivial transformations one can easily single out the factor
$(1+tr(a))$ in (\ref{b}) and get

\begin{equation}
E^{\alpha}E_{\alpha}(\Phi(1+tr(a)) + c^{\alpha \enspace \sigma}_{\enspace 
\sigma}E_{\alpha}(\Phi(1+tr(a))) +
\Phi(1+tr(a)){\cal V}(c)=\Box_{E}(\Phi(1+tr(a))) 
\end{equation} 
 
The last equation means that the right hand side of (\ref{mainformula})
satisfies the infinitesimal version of (\ref{consq}),
so that by the above considerations it really represents the operation
$\Box_{E}$.

%%%%%%%%%%%%%%%%%%%%%%%%%%%%%%%%%%%%%%%%%%%%%%%%%%%%%%%%%%%%%
\section{Construction of the Lagrangian}
We start with the term describing the interaction between matter chiral fields
$\Phi_{i}$. It has the form
\begin{equation} \label{pot}
\int dV {\cal U}(\Phi) + h.c. =
\int dV\left[a_{i}\Phi_{i}+\frac{1}{2}m_{ij}\Phi_{i}
\Phi_{j}+\frac{1}{3}g_{ijk}\Phi_{i}\Phi_{j}\Phi_{k}
\right] + h.c.
\end{equation}
where the function ${\cal U}(\Phi)$ is assumed to be gauge invariant.
The integration here is as described in section 3. 
As we do not have any metric, we have to restrict ourselves to
SCR-transformations in order to preserve the volume element $dV$.

%%%%%%%%%%%%%%%%%%%%%%%%%%%%%%%%%%%
%%%%%%%%%%%%%%%%%       KINETIC TERM
%%%%%%%%%%%%%%%%%%%%%%%%%%%%%%%%%%%%%%%%
To construct the kinetic term one needs to apply first a chiral projector $\Box$
to  the gauge invariant quantity $\Phi^{+}e^{V}\Phi$ and then perform the
integration described above. The explicit construction of a chiral projector in
terms of internal geometry of the surface $\Omega$ was given in section 2. It was
shown there that the action of a generic chiral projector can be written as follows
$$
\bar \Box_{\bar E}\rho \Phi^{+}e^{V}\Phi
$$
where $\bar \Box_{\bar E}$ is a chiral projector corresponding to a natural
volume element defined by the basis fields $\bar E_{\dot \alpha}$,
the explicit construction of which is given by formula (\ref{mainformula}),
and $\rho$ is an arbitrary function. Note that under a change of basis, 
the vector fields $\bar E_{\dot \alpha}$
corresponding to (\ref{scrtr}) the operation $\bar \Box_{\bar E}$ behave as follows
$$ 
\bar \Box_{\bar E'}=\Box_{\bar E}\det(\bar g)
$$
(see also formulae (\ref{transform}) and (\ref{consq}) in  appendix A).
Therefore to have SCR covariance one has to chose the transformation law for
$\rho$ to be $\rho'=\frac{\rho}{\det\bar g}$. It can be easily
checked that $\rho=|\det\Gamma|^{-\frac{1}{3}}$ has this property. 
Here $|\det\Gamma|$ stands for the  absolute value
of the determinant of  the Levi form corresponding to the surface $\Omega$. 
Therefore the SCR covariant expression of the kinetic term reads as
\begin{equation} \label{kinetic}
\int dV\left[\bar \Box_{\bar E}\left| \frac{1}{4}\det\Gamma 
\right|^{-\frac{1}{3}}\Phi^{+}e^{V}\Phi\right] + h.c.
\end{equation} 
where $\frac{1}{4}$ is inserted for normalization purposes.  

%%%%%%%%GAUGE LAGRANGIAN
%%%%%%%%%%%%%%%%%%%%%%%%%%%%%%%%%%%%%%%%
Finally let us turn to the construction of the Lagrangian of gauge fields.
This turns out to be the most technically complicated part of the whole
construction. First note that due to our large gauge invariance group,
(\ref{gaugefi}) when $\Lambda$ and $\bar \Lambda$ are arbitrary functions with
values in the gauge Lie algebra it is impossible to construct the gauge strength
fields $W_{\alpha}, W_{\dot \alpha}$ corresponding to each of fields
$e^{V}, e^{-U}, e^{\tilde U}$.  Consider the combination
$e^{G}=e^{\tilde U}e^{-V}e^{-U^{t}}$, which is invariant under gauge
transformations. Due to the 
arbitrariness in the choice  of solutions $U$ and $\tilde U$ 
to the zero curvature equations (\ref{zerocurv}) mentioned in section 3, 
there is an ambiguity 
in the definition of $e^{G}$ described by the transformation
\begin{equation} \label{trG}
e^{G'}=e^{T}e^{G}e^{(S^{+})^{t}}
\end{equation}
where $T$ and $S^{+}$ are arbitrary respectively chiral and antichiral
$\cal G$-valued functions.

The generic form of chiral field strengths for $e^{G}$ reads as

\begin{equation} \label{W}
 W_{\alpha}=\bar \Box_{\bar E}B_{\alpha}^{\beta}e^{G}E_{\beta}e^{-G}
\end{equation}
where $B_{\alpha}^{\beta}$ is a matrix to be defined by the invariance
properties of $W_{\alpha}$.
The transformation law for $B_{\alpha}^{\beta}$ under (\ref{scrtr}) can be
derived in terms of the invariance of construction (\ref{W}) under (\ref{scrtr}),
and has the form

\begin{equation} \label{trB}
(B')_{\alpha}^{\beta}=
\frac{1}{\det(\bar g)}B_{\alpha}^{\gamma}(g^{-1})^{\beta}_{\gamma}
\end{equation}
Further restrictions on the matrix $B$ come from the covariance requirement
under the substitution (\ref{trG}). For $W_{\alpha}$ to transform covariantly
under (\ref{trG}), the operator $\bar \Box_{\bar E}B_{\alpha}^{\beta}E_{\beta}$
ought to annihilate chiral functions (the annihilation of
antichiral functions is obvious).  It is convenient to derive the restriction
on the matrix $B$ first in a special basis, namely where 
$\{\bar E_{\dot \alpha},\bar E_{\dot \beta}\}=0$ (see appendix A for details). 
But first let us introduce some useful notations. One can choose the vectors 
$E_{\alpha \dot \beta}=\{E_{\alpha},\bar E_{\dot \beta}\}$ as a basis of the 
real tangent space to $\Omega$.  Then we have the following commutation
relations

\begin{equation} \label{crel}
[\bar E_{\dot \gamma}, E_{\alpha \dot \beta} ]=
{\check c}_{\gamma, \alpha \dot \beta}^{\enspace
\enspace \enspace \, \sigma \dot \sigma}
E_{\sigma \dot \sigma} + 
{\check c}_{\gamma, \alpha \dot \beta}^{\enspace \enspace \enspace
\, \sigma}E_{\sigma}  + {\check c}_{\gamma, \alpha \dot \beta}^{\enspace
\enspace \enspace \, \dot \sigma}\bar E_{\dot \sigma}
\end{equation}
where 
${\check c}_{\gamma, \alpha \dot \beta}^{\enspace \enspace \enspace \, A}$ 
are some functions.
The connection between the two bases introduced in real space is given by
the following formulas

\begin{eqnarray}
E_{\alpha \dot \beta}&=&{c_{\alpha\dot \beta}}^{a}E_{a} +
{c_{\alpha\dot \beta}}^{\gamma}E_{\gamma} +
 {c_{\alpha\dot \beta}}^{\dot \gamma}\bar E_{\dot \gamma} \nonumber \\
E_{a}&=&\bar {c_{a}}^{\alpha \dot \beta}E_{\alpha \dot \beta} -
\bar {c_{a}}^{\gamma}E_{\gamma} - 
\bar {c_{a}}^{\dot \gamma}\bar E_{\dot \gamma}            \label{twobas}
\end{eqnarray}
where 
$\bar {c_{a}}^{A}=i(\Gamma^{-1})_{a}^{b}\bar \sigma^{\alpha \dot
\beta}_{b}{c_{\alpha \dot \beta}}^{A}$.
We use the notation 
$\bar \sigma^{\alpha \dot \beta}_{b}$ for the Pauli matrices and 
$\Gamma^{-1}$ is the inverse of the Levi form matrix 
$\Gamma_{a}^{b}=i\bar \sigma_{a}^{\alpha
\dot \beta}{c_{\alpha \dot \beta}}^{b}$.
Substituting the first expression (\ref{twobas}) into the commutator on the
left hand side of (\ref{crel}),
one obtains the following expression for the quantity 
${\check c}_{\gamma, \sigma \dot \beta}^{\enspace \enspace \enspace \,
\sigma \dot \sigma}$ in terms of the functions ${c_{AB}}^{D}$

\begin{equation} \label{a}
{\check c}_{\gamma, \sigma \dot \beta}^{\enspace \enspace \enspace \,
\sigma \dot \sigma}=
\bar {c_{a}}^{\sigma \dot \sigma}E_{\gamma}{c_{\sigma \dot \beta}}^{a} +
{c_{\sigma \dot \beta}}^{a}\bar {c_{b}}^{\sigma \dot \sigma}
{c_{\gamma a}}^{b} - {c_{\gamma \dot \beta}}^{\dot \sigma}
\end{equation}
Contracting the upper index $\dot \sigma$ with the lower $\dot \beta$
we obtain the following important identity which we will need later

\begin{equation} \label{important}
{{\check c}_{\gamma, \sigma \dot \sigma}}^{\enspace \enspace \enspace \,
\sigma \dot \sigma} = 
\bar {c_{a}}^{\sigma \dot \sigma}E_{\gamma}{c_{\sigma \dot \sigma}}^{a} + 
({c_{\gamma a}}^{a} - 
{c_{\gamma \dot \sigma}}^{\dot \sigma})=(\det\Gamma)^{-1}E_{\gamma}\det\Gamma
\end{equation}
where in the last step we used the identity (\ref{scrid}).
The complex conjugate to equation (\ref{important}) reads as
\begin{equation} \label{conjimp}
{{\check c}_{\dot \gamma, \dot \sigma  \sigma}}^{\enspace \enspace \enspace
\, \dot \sigma  \sigma} = 
(\det\Gamma)^{-1}\bar E_{\dot \gamma}\det\Gamma
\end{equation}

Now we are ready to explore the restrictions on the matrix $B$ in (\ref{W}).
For the basis specified above, applying the operator 
$\Box_{\bar E} B_{\alpha}^{\beta}E_{\beta}$ to some chiral function $\Phi$
we get
\begin{eqnarray}  \label{source}
\Box_{\bar E} B_{\alpha}^{\beta}E_{\beta}\Phi = 
 \bar E^{\dot \gamma}\bar E_{\dot \gamma}B_{\alpha}^{\beta}E_{\beta}\Phi = 
[\bar E^{\dot \gamma}, \{\bar E_{\dot \gamma},
B_{\alpha}^{\beta}E_{\beta}\}]\Phi = \nonumber \\ 
2B_{\alpha}^{\delta}\epsilon^{\dot \sigma \dot \delta}
((B^{-1})^{\gamma}_{\delta}\bar E_{\dot \delta}B_{\gamma}^{\sigma} + 
{\check c}_{\dot \delta , \dot \sigma \delta}^{\enspace \enspace \enspace \,
\dot \sigma \sigma})
E_{\sigma \dot \sigma}\Phi  + f_{\alpha}^{\beta}E_{\beta}\Phi
\end{eqnarray} 
where by $f_{\alpha}^{\beta}$ we denote some functions which can be expressed
in terms of the matrix $B$ and its derivatives, but are unessential for further
analysis.  The last remark is due to the fact that if the coefficient 
of $E_{\sigma \dot \sigma}$ vanishes, then the whole operator vanishes
on chiral functions.  This is because the operator on the left hand side of
(\ref{source}) acts from chiral functions to chiral ones (for the fixed index
$\alpha$) and it can be easily shown that an operator with this property and of
the form $f_{\alpha}^{\beta}E_{\beta}$ is identically zero.
Therefore we have the following equation for the matrix $B$

\begin{equation} \label{Beq}
(B^{-1})_{\delta}^{\gamma}\bar E_{\dot \delta}B_{\gamma}^{\sigma} = 
- {\check c}_{\dot \delta , \dot \sigma \delta}^{\enspace
\enspace \enspace \, \dot \sigma \sigma}
\end{equation}
%It can be viewed as a claim that the coefficients 
% $- {\check c}_{\dot \delta , \dot \sigma \delta}^{\enspace \enspace \enspace \,
%\dot \sigma \sigma}$ 
%correspond to coefficients of a plain semiconnection responsible for the
%parallel transport of complex
%tangent vectors in $\dot E_{\dot \alpha}$ directions.
It can be derived directly from the definition (\ref{crel}) that 
the transformation law for the coefficients 
$- {\check c}_{\dot \delta , \dot \sigma \delta}^{\enspace \enspace \enspace \,
\dot \sigma \sigma}$ under (\ref{scrtr}) is as follows
\begin{equation} \label{trcheckc}
- ({\check c \, '})_{\dot \delta , \dot \sigma \delta}^{\enspace \enspace
\enspace  \dot \sigma \sigma} = 
-{\check c}_{\dot \nu , \dot \sigma \gamma}^{\enspace \enspace \enspace \,
\dot \sigma \beta}   
\bar g^{\dot \nu}_{\dot \delta}g^{\gamma}_{\delta}(g^{-1})^{\sigma}_{\beta} + 
\bar g^{\dot \nu}_{\dot \delta}g^{\gamma}_{\delta}\bar E_{\dot \nu}
(g^{-1})_{\gamma}^{\sigma} - 
\delta^{\sigma}_{\delta}\bar g^{\dot \nu}_{\dot \delta}
(\det\bar g)^{-1}\bar E_{\dot \nu}\det\bar g
\end{equation}
where $\delta^{\sigma}_{\delta}$ is the Kronecker symbol.
The first two terms in this expression are of the same form as the
transformation law for semiconnection coefficients.
One can easily check  using (\ref{trB}) that the transformation rule for the
expression on the left hand side of (\ref{Beq}) is of the same form,
which is of course what one should expect since  the operator 
under consideration is defined invariantly. 
This invariance implies that it is
sufficient to prove the solvability of (\ref{Beq}) in the special basis which
we have already used. In this basis the integrability condition for  (\ref{Beq})
takes the form
\begin{equation} \label{intcond}
\bar E_{\dot \alpha}{\check c}_{\dot \beta ,
\dot \sigma \delta}^{\enspace \enspace \enspace \, \dot \sigma \sigma}  
+ \bar E_{\dot \beta}
{\check c}_{\dot \alpha, \dot \sigma \delta}^{\enspace \enspace \enspace \,
\dot \sigma \sigma} - 
{\check c}_{\dot \alpha, \dot \sigma \delta}^{\enspace \enspace \enspace \,
\dot \sigma \gamma} {\check c}_{\dot \beta, \dot \nu \gamma}^{\enspace
\enspace \enspace \, \dot \nu \sigma} - 
{\check c}_{\dot \beta , \dot \sigma \delta}^{\enspace \enspace \enspace \,
\dot \sigma \gamma} {\check c}_{\dot \alpha , \dot \nu \gamma}^{\enspace
\enspace \enspace \, \dot \nu \sigma} = 0
\end{equation}
This formula follows from two identities
\begin{eqnarray} 
&& \{ \bar E_{\dot \alpha}, [ \bar E_{\dot \beta}, E_{\gamma \dot \gamma}
] \} = 0    \nonumber  \\
  &&   [     {\bar E}_{\dot \beta}   ,   E_{\gamma \dot \alpha}      ]    +      
   [ \bar E_{\dot \alpha}, E_{\gamma \dot \beta} ]  = 0    \label{identities}
\end{eqnarray}
which are true when $\{\bar E_{\dot \alpha}, \bar E_{\dot \beta}\}=0$. 
The second identity in (\ref{identities}) can be equivalently written as 
$$
{\check c}_{\dot \alpha, \dot \beta \delta}^{\enspace \enspace \enspace
\, A} + 
{\check c}_{\dot \beta, \dot \alpha \delta}^{\enspace \enspace \enspace
\, A} = 0 
$$
Writing the first identity in (\ref{identities}) in terms of the basis 
$E_{\alpha}, \bar E_{\dot \alpha}, E_{\alpha \dot \alpha}$, for the coefficient of 
$E_{\sigma \dot \sigma}$ we have 
$$
- {\check c}_{\dot \beta, \dot \gamma \gamma}^{\enspace
\enspace \enspace \, \dot \mu \nu}
{\check c}_{\dot \alpha, \dot \mu \nu}^{\enspace \enspace \enspace \, \dot
\sigma \sigma} + \bar E_{\dot \alpha}
{\check c}_{\dot \beta, \dot \gamma \gamma}^{\enspace \enspace \enspace \,
\dot \sigma \sigma} + \delta^{\dot \sigma}_{\dot \alpha}
{\check c}_{\dot \beta, \dot \gamma \gamma}^{\enspace \enspace \enspace \, 
\sigma} = 0
$$
Contracting the indices $\dot \sigma$ and $\dot \gamma$ in the last expression,
symmetrizing  it over the indices $\dot \alpha, \dot \beta$ and using the second
identity in (\ref{identities}) we get  (\ref{intcond}). 
To summarize, we showed that equation (\ref{Beq}) is solvable. 
Note however that there is a large arbitrariness in the choice of solutions to
(\ref{Beq}). Namely, one can multiply any solution by a matrix with chiral
entries and get another one. We will show how to narrow the 
choice of solution. But first let us write the Lagrangian 
$W^{\alpha}W_{\alpha}$ in terms of the construction
(\ref{W}) taking into account (\ref{Beq}).
After some trivial transformations we will get
\begin{equation} \label{glag}
W^{\alpha}W_{\alpha} = \det B(\Box_{\bar{\cal D}}e^{G}E^{\alpha}e^{-G})
(\Box_{\bar {\cal D}}e^{G}E_{\alpha}e^{-G})
\end{equation}    
where $\Box_{\bar {\cal D}}$ is obtained from the usual chiral
projector $\Box_{\bar E}$ by substitution
of the derivatives $\bar E_{\dot \alpha}$ by covariant derivatives
$\bar {\cal D}_{\dot \alpha}$
which act on a tensor $V$ carrying a spinor index $\alpha$ in the following way
$$
{\bar {\cal D}_{\dot \alpha}}V_{\beta} = \bar E_{\dot \alpha}V_{\beta} - 
{\check c}_{\dot \alpha, \dot \sigma \beta}^{\enspace \enspace \enspace \,
\dot \sigma \sigma} V_{\sigma}
$$
The only quantity in (\ref{glag}) which depends on the choice of the solution to
(\ref{Beq}) is $\det B$.  From (\ref{Beq}) we have
$$
\det B\bar E_{\gamma}(\det B)^{-1} = 
{\check c}_{\dot \gamma, \dot \sigma
\sigma}^{\enspace \enspace \enspace \, \dot \sigma \sigma}
$$
As one can see from (\ref{conjimp}),  the determinant of the Levi form $\Gamma$
satisfies exactly the same equation. 
Therefore one can take $\det\Gamma$ multiplied by a constant factor as a solution
to the last equation.
We choose $\det B=4(\det\Gamma)^{-1}$. With this choice of constant factor we
will get the standard form 
for our Lagrangian  in a Wess-Zumino gauge for the basis vector fields.

%%%%%%%%%%%%%%%%%%%%%%%%%%%%%%%%%%%%%%%%%%%%%%
%%%%%%%%%%%%%    Acknowledgment
{\bf Acknowledgements}:
We are indebted to D. Kazhdan for interesting discussions.  
We also want to express our gratitude to M. Penkava for reading 
and editing the manuscript. 

%%%%%%%%%%%%%%%%%%%%%%%%%%%%%%%%%%%%%%%%%%%%%%%%%%%%%%%%%

%%%%%%%%%%%%%%%%%%%%%%%%%%%%%%%%%%%%%%%%%%%%%%%%%%%
%%%%%%%%%%%%%     REFERENCES         %%%%%%%%%%%%%%%%%%%%%%%%%
%%%%%%%%%%%%%%%%%%%%%%%%%%%%%%%%%%%%%%%%%%%%%%%%%%%

 \end{document}